\documentclass[10pt]{revtex4}

\usepackage{amssymb,amsmath,amsfonts}
\usepackage{graphicx,graphics}
\usepackage{color}

\newcommand{\rem}[1]{}
\DeclareMathAlphabet{\mathbi}{OML}{cmm}{b}{it}
\newcommand{\non}{\nonumber}
\newtheorem{theorem}{Theorem}

\newtheorem{corollary}{Corollary}

\newcommand{\bx}{\mathbi{x}}

\newcommand{\bel}{\begin{equation}\label}
\newcommand{\ee}{\end{equation}}
\newcommand{\beq}{\begin{eqnarray}\label}
\newcommand{\eeq}{\end{eqnarray}}
\newcommand{\bc}{\begin{center}}
\newcommand{\ec}{\end{center}}
\newcommand{\ben}{\begin{enumerate}}
\newcommand{\een}{\end{enumerate}}
\newcommand{\bit}{\begin{itemize}}
\newcommand{\eit}{\end{itemize}}
\newcommand{\I}{\int_{\Omega}}
\newcommand{\bB}{\mbox{\boldmath$B$}}
\newcommand{\bhatB}{\mathbi{\hat{B}}}

\newcommand{\bhn}{\mathbi{\hat{n}}}

\newcommand\shalf{\ensuremath{{\scriptstyle\frac{1}{2}}}}

\newcommand{\bu}{\mbox{\boldmath$u$}}
\newcommand{\bv}{\mathbi{v}}

\newcommand{\bom}{\mbox{\boldmath$\omega$}}

\newcommand{\bk}{\mbox{\boldmath$\hat{k}$}}

\begin{document}

\title{\textbf{The 3D incompressible Euler equations with a passive scalar\,:\\
a road to blow-up?}}

\author{John D. Gibbon$^{\hbox{\scriptsize a}}$}
\affiliation{$^{\hbox{\scriptsize\itshape a}}$Department of Mathematics, Imperial College London SW7 2AZ, UK.}
\email{j.d.gibbon@ic.ac.uk}
\author{Edriss S. Titi$^{\hbox{\scriptsize b,\,c}}$}
\affiliation{$^{\hbox{\scriptsize\itshape b}}$Department of Computer Science and Applied Mathematics, Weizmann Institute of
Science, 76100 Rehovot, Israel\,,\\}
\affiliation{$^{\hbox{\scriptsize\itshape c}}$ Department of Mathematics and Department of Mechanical and Aerospace Engineering,
University of California, Irvine, California 92697, USA.}
\email{etiti@math.uci.edu}

\begin{abstract}
The  3D incompressible Euler equations with a passive scalar $\theta$ are considered in a smooth domain $\Omega\subset \mathbb{R}^{3}$
with no-normal-flow boundary conditions $\bu\cdot\bhn|_{\partial\Omega} = 0$. It is shown that smooth solutions blow up in a finite time if
a null (zero) point develops in the vector $\bB = \nabla q\times\nabla\theta$, provided $\bB$ has no null points initially\,:  $\bom =
\mbox{curl}\,\bu$ is the vorticity and $q = \bom\cdot\nabla\theta$ is a potential vorticity. The presence of the passive scalar concentration
$\theta$ is an essential component of this criterion in detecting the formation of a singularity. The problem is discussed in the light of a 
kinematic result by Graham and Henyey (2000) on the non-existence of Clebsch potentials in the neighbourhood of null points.
\end{abstract}

\maketitle
\pagestyle{myheadings}\markright{1st/06/13 \hfil 3D incompressible Euler equations with a passive scalar\,: a road to blow up?}

\par\vspace{-5mm}

\section{Introduction}

It is known that the 3D incompressible Euler equations have short time well-posdeness of strong solutions (see, e.g., Ebin 
and Marsden 1970, Kato 1972, Kato and Lai 1984, Lichtenstein 1925--1930, Temam 1975); and an array of very weak solutions 
(see Scheffer 1993, Shnirelman 1997, Brenier 1999, Bardos and Titi 2007,\,2010,\,2013, De Lellis and Sz\'ekelyhidi 2009,
\,2010, Wiedemann 2011, Bardos, Titi and Wiedemann 2012), but whether a singularity develops from smooth initial 
conditions in a finite time has been a controversial open problem for a generation (Beale, Kato and Majda 1984,  Kerr 1993, 
Constantin, Fefferman and Majda 1996, Hou and Li 2006, Constantin 2007,  Bustamante and Kerr 2008, Gr\"afke, Homann, 
Dreher and Grauer 2008, Hou 2008). Most numerical experiments are performed on periodic boundary conditions\,: more 
than twenty of these are cited in the review by Gibbon (2008). In contrast, the aim of this paper is to study the blow-up 
problem in the context of the evolution of divergence-free solutions of the Euler equations $\bu(\bx,t)$ together with a 
passive scalar $\theta(\bx,t)$
\bel{eul1a}
\frac{D\bu}{Dt} = - \nabla p\,,\qquad \frac{D\theta}{Dt} = 0\,,\qquad
\frac{D~}{Dt} = \partial_{\,t} + \bu\cdot\nabla\,,
\qquad\mbox{div}\,\bu = 0\,,
\ee
in a smooth finite domain $\Omega \subset \mathbb{R}^{3}$ with no-normal-flow boundary conditions $\bu\cdot\bhn|_{\partial\Omega} =0$.
The inclusion of $\theta$, which could represent any passive tracer concentration (Constantin 1994, Constantin, Procaccia and Sreenivasan 1991,
Constantin and Procaccia 1993), allows the vector $\nabla\theta$ to interact with the fluid vorticity field $\bom = \mbox{curl}\,\bu$ which evolves
according to
\bel{eulvort1}
\frac{D\bom}{Dt} = \bom\cdot\nabla\bu\,.
\ee
Formally, it is easily shown that the equivalent of potential vorticity $q=\bom\cdot\nabla\theta$ is also a passive quantity\,: see Hoskins, McIntyre
and Robertson (1985) for a more general discussion of potential vorticity in the geophysical context. The result is related to what is
known as Ertel's Theorem (Ertel 1942) which is derived via a simple geometric expression
\beq{q1A}
\frac{Dq}{Dt} = \left(\frac{D\bom}{Dt} - \bom\cdot\nabla\bu \right)\cdot\nabla \theta
+ \bom\cdot\nabla\left(\frac{D\theta}{Dt}\right)\,.
\eeq
This is no more than a re-arrangement of terms after an application of the product rule. Clearly $q$ satisfies
\bel{q1B}
\frac{Dq}{Dt} = 0\,.
\ee
A result of Moiseev, Sagdeev, Tur and Yanovskii  (1982) (see also  Kurgansky and Tatarskaya (1987), Yahalom (1996), 
Kurgansky and Pisnichenko (2000), Gibbon and Holm 2010,\,2012) can now be invoked for any two passive scalars 
whose gradients define the vector
\bel{Bdef}
\bB = \nabla q\times\nabla\theta\,,
\ee
in which case $\bB$ turns out to satisfy
\bel{Bev1}
\frac{D\bB}{Dt} = \bB\cdot\nabla\bu\,,\qquad\qquad\mbox{div}\,\bB =0\,,\qquad\qquad\mbox{div}\,\bu =0\,.
\ee
$\bB$ is the cross-product of the two normals to the material surfaces on which $\theta$ and $q$ evolve and is thus tangent to the
curve formed from the intersection of the two surfaces (Gibbon and Holm 2010,\,2012). In our case, with $q=\bom\cdot\nabla\theta$,
the vector $\bB$ contains the \textit{gradient} of $\bom$ (in projection) and two gradients of $\theta$. $\bB$ is a construction or 
`Clebsch representation' of the vector $\bB$ in terms of the `potentials' $(q,\,\theta)$ rather than a decomposition.
Nevertheless, we propose to exploit the fact that the evolution of $\bB$ in (\ref{Bev1}) takes the same form as that of the 
Euler vorticity field in (\ref{eulvort1}) or of a magnetic field in MHD (Moffatt 1969). For general divergence-free vector fields, Arnold 
(1978) and Arnold and Khesin (1998) remarked that such flows under a Clebsch decomposition are simple and that tangency of one 
level set of a Clebsch potential with another (forming a null point) is not allowed. This was reinforced by a kinematic result of Graham 
and Henyey (2000) on the non-existence of these potentials in a neighbourhood that contains a null point.  Putting this in the context 
of the results in this current paper, any momentary alignment of $\nabla q$ and $\nabla\theta$ at a time $t^{*}$ causes a null point 
in the $\bB$-field\,: with $q$ and $\theta$ passively transported by a 3D Euler flow, this leads to the breakdown of regularity of the 
underlying Euler velocity field at $t^{*}$.


\section{Statement of the result}\label{psresult}

The Beale-Kato-Majda (BKM) theorem is the main regularity result for the 3D Euler equations (Beale, Kato and Majda 1984). In its original 
form it was proved in the whole of $\mathbb{R}^3$ and used the $L^{\infty}$-norm of the vorticity, $\|\bom\|_{L^{\infty}(\mathbb{R}^{3})}$ 
as its key object. A subsequent simple modification was proved by Ponce (1985) in terms of the rate of strain matrix (deformation tensor)
defined by $\mathcal{S} = \shalf\left(\nabla\bu + \nabla\bu^{T}\right)$\,:
\begin{theorem}\label{ponce}
There exists a global strong solution of the 3D Euler equations,
$\bu \in C([0,\,\infty);H^{s})\cap C^{1}([0,\,\infty);H^{s-1})$ for $s\geq 3$ , in $\mathbb{R}^3,$ if and only if  for every $t>0$,
\bel{BKM1}
\int_{0}^{t}\|\mathcal{S}(\tau)\|_{L^{\infty}(\mathbb{R}^3)}\,d\tau <\infty\,.
\ee
Equivalently, 
\bel{BKM2} 
\int_{0}^{t}\|\mathcal{S}(\tau)\|_{L^{\infty}(\mathbb{R}^3)}\,d\tau \to \infty\,,\qquad\mbox{as}\qquad t \nearrow t^{*},
\ee
for some $t^{*}>0$, if and only if  the corresponding strong solution of the three-dimensional Euler equations, with initial data in $H^s(\mathbb{R}^3)$, for $s\geq 3$, blows up within the interval $[0,t^{*}]$.
\end{theorem}
The proofs in Beale \textit{et al.} (1984) and Ponce (1985) are valid for flow in all $\mathbb{R}^{3}$, but the techniques used in those papers, such as 
Fourier transforms and the Biot-Savart integral, are not readily adaptable for ideal flows in a bounded smooth domain $\Omega$ subject to no-normal-flow boundary conditions $\bu\cdot\bhn|_{\partial\Omega} = 0$. These difficulties were circumvented by  Ferrari (1993) and Shirota and Yanagisawa (1993) 
who adapted some ideas from the theory of linear elliptic systems to achieve a proof of the BKM theorem for ideal flows in bounded smooth domains, 
subject to the no-normal-flow boundary conditions $\bu\cdot\bhn|_{\partial\Omega} = 0$. A further simple adaptation of these methods allows us prove 
the analogue of Theorem \ref{ponce} for the three-dimensional Euler equations in smooth bounded domains, subject to the boundary conditions 
$\bu\cdot\bhn|_{\partial\Omega} = 0$. For our purposes, this modification of (\ref{BKM2}) is the key blow-up result for the bounded smooth domain $\Omega$.
\par\smallskip
The main questions revolve around the occurrence of null points (zeros) in $|\bB(\bx,\,t)|$. Firstly, initial data for $|\bB|$ 
must be free of null points. Given the definition of $\bB$ this means initial data must be free of maxima or minima in $\theta$ or $q$, 
nor must $\nabla q$ and $\nabla\theta$ be aligned or anti-aligned at any point. Since $q$ and $\theta$ are passively transported, the 
flow will be free of maxima and minima in these variables for all time. Thus the only way a null can develop for $t>0$ is through a 
momentary alignment or anti-alignment of $\nabla q$ with $\nabla\theta$. \S\ref{id} contains an example of simple initial data and 
a domain $\Omega$ with no null points for $|\bB|$. In the following, $t^{*}$ is designated as the earliest time a null point occurs in 
$|\bB(\bx,\,t)|$.
\begin{theorem}\label{Bthm}
Let $\bu$ be a strong solution of the  three-dimensional Euler equations in a bounded smooth domain $\Omega\subset \mathbb{R}^{3}$, 
with boundary conditions $\bu\cdot\bhn|_{\partial\Omega} = 0$. Let $\bB$ be the solution of (\ref{Bev1}) with initial data for which 
$|\bB(\bx,\,0)| > 0$ and $\|\bB\left(\cdot\,,0\right)\|_{L^{\infty}(\Omega)} < \infty$. Assume that the solution $\bu$ exists on  the interval 
$[0,\,t^{*})$, where $t^{*}$ the earliest time at which $\lim_{t \nearrow t^{*}}|\bB(\bx,\,t)| = 0$, for some $\bx \in \Omega$, then
\bel{thm1A} 
\int_{0}^{t}\|\mathcal{S}(\tau)\|_{L^{\infty}(\Omega)}\,d\tau\to \infty\,,\qquad\mbox{as}\qquad t \nearrow t^{*}\,,
\ee
i.e., $t^{*}$ is a blow up time of the solution $\bu$. 
Equivalently, if $\bu$ is a strong solution of three-dimensional Euler equations on the interval [0,T], for some $T>0$,  (in particular we
 have $\int_{0}^{T}\|\mathcal{S}(\tau)\|_{L^{\infty}(\Omega)}\,d\tau < \infty$) then for every    $t\in [0,\,T]$  $|\bB(\bx,\,t)|\neq 0$.
\end{theorem}

\par\smallskip\noindent
\textbf{Proof\,:} On the interval $[0,\,t^{*})$ first take the scalar product of (\ref{Bdef}) with $\bB$,
divide by $B^{2} = \bB\cdot\bB$, and then multiply by $(\ln B) |\ln B|^{2(m-1)}$ (which could take either sign) to obtain
(for $1 \leq m < \infty$)
\bel{b4}
\frac{1}{2m}\frac{D|\ln B|^{2m}}{Dt} = (\ln B)|\ln B|^{2(m-1)}\left(\bhatB\cdot \mathcal{S}\cdot\bhatB\right)\,.
\ee
Next integrate over the volume, invoke the Divergence Theorem and the boundary conditions on $\Omega$ and finally use
H\"older's inequality to obtain
\beq{b5}
\frac{1}{2m}\frac{d~}{dt}\I |\ln B|^{2m}dV &\leq&
\I \left|\ln B\right|^{2m-1}\left|\mathcal{S}\right|\,dV\non\\
&\leq& \left(\I |\ln B|^{2m}dV\right)^{\frac{2m-1}{2m}}\left(\I |\mathcal{S}|^{2m}dV\right)^{\frac{1}{2m}}\,.
\eeq
Using the standard notation $\|X\|_{L^{p}(\Omega)} = \left(\I |X|^{p}dV\right)^{1/p}$, (\ref{b5}) reduces to
\bel{b6}
\frac{d~}{dt}\|\ln B\|_{L^{2m}(\Omega)} \leq \|\mathcal{S}\|_{L^{2m}(\Omega)}\,,
\ee
which integrates to
\bel{b7B}
\|\ln B(t)\|_{L^{\infty}(\Omega)} \leq \|\ln B(0)\|_{L^{\infty}(\Omega)} + \int_{0}^{t}\|\mathcal{S}(\tau)\|_{L^{\infty}(\Omega)}\,d\tau,
\ee
in the limit $m\to\infty$ ($\Omega$ is bounded).
Provided $\bB$ has no zero in its initial data, the log-singularity at $|\bB| = 0$ causes the left hand side to blow up at $t^{*}$ thereby
forcing $\int_{0}^{t}\|\mathcal{S}\|_{L^{\infty}(\Omega)}\,d\tau \to \infty$ as $t\nearrow t^{*}$. This result could also be obtained 
using Lagrangian particle paths if one choose that route to perform the analysis. Finally, it is immediately clear from (\ref{b7B}) that 
if $\int_{0}^{T}\|\mathcal{S}\|_{L^{\infty}(\Omega)}\,d\tau$ is finite, then no null can develop in $\bB(\bx,\,t)$, for  $t \in [0,\,T]$.
\hfill $\blacksquare$
\par\medskip\noindent
\textbf{Remark\,:} The fate of $q$ and $\theta$ as $t\nearrow t^{*}$ is answered by a purely kinematic result of Graham and 
Henyey (2000) which states that Clebsch potentials do not exist in some neighbourhood, however small, containing a generic null 
point. A result that is consistent with Theorem \ref{Bthm} follows as a direct consequence\,:
\begin{corollary}
Assume that initially $(q,\,\theta) \in L^{\infty}\left(\Omega\right)$ and that $\bB$ develops a null point at $t = t^{*}$.
The kinematic result of Graham and Henyey (2000) implies that it is impossible to find potentials $(q,\,\theta)$ in some
neighbourhood of this null point\,: that is, as $t\nearrow t^{*}$ the variables $(q,\,\theta) \notin L^{\infty}\left(\Omega\right)$
in this neighbourhood. Thus the advecting velocity field $\bu$ transporting $q$ and $\theta$ must become non-smooth
in this limit.
\end{corollary}

\section{An example of initial data with no null points}\label{id}

We proceed to find a simple example of a set of initial data $\bu$ on a finite domain $\Omega \subset \mathbb{R}^{3}$ from initial
data on $\bom$ and $\theta$ such that $|\bB|>0$ and $\|\bB\|_{L^\infty(\Omega)} < \infty$  for the elliptic system
\bel{id1}
\mbox{curl}\,\bu = \bom\,,\qquad\qquad\mbox{div}\,\bu =0\,,\qquad\qquad
\bu\cdot\bhn|_{\partial\Omega} =0\,.
\ee
The usual methods, such as the Biot-Savart integral, are hard to apply on this domain, but for the elliptic system in (\ref{id1}), it can 
be shown that for given a vector $\bom$, the velocity field $\bu$ can, in principle, always be constructed (see Ferrari 1993 and
Shirota and Yanagisawa 1993). In the next paragraph this construction is performed in an explicit example in which $\Omega$
will be determined later.
\par\smallskip
Take the example $\bom =(1,\,1,\,1)^{T}$\,: we firstly observe that there is a velocity field $\bv = (z,\,x,\,y)^{T}$ which
satisfies $\mbox{div}\,\bv =0$ and $\mbox{curl}\,\bv = (1,1,1)^{T}$, but we cannot be sure that it satisfies $\bv\cdot\bhn = 0$
for any given domain $\Omega$. Therefore it needs to be modified to satisfy the boundary conditions.
To do this we introduce some potential $\phi$ such that
\bel{id2}
\bu = \bv + \nabla \phi\,.
\ee
Note that $\mbox{curl}\,\bu = (1,\,1,\,1)^{T}$. To guarantee that (\ref{id1}) holds, $\phi$ must satisfy the Neumann
boundary value problem
\bel{id3}
\Delta\phi = 0\,,\qquad\qquad
\left. \frac{\partial\phi}{\partial n}\right|_{\partial\Omega} = - (z,\,x,\,y)^{T}\cdot\bhn\,,
\ee
which always has a solution on any smooth domain $\Omega$. Thus we have been able to construct a velocity field
$\bu$ corresponding to $\bom = (1,\,1,\,1)^T$, that satisfies the boundary conditions. For simplicity, now choose
$\theta = \shalf(x^{2}+y^{2}+z^{2})$ (say) and calculate $q$, $\nabla q$ and $\nabla\theta$
\bel{id2}
q = x+y+z\,,\qquad\nabla\theta = (x,\,y\,,z)^{T}\,,\qquad\nabla q = (1,\,1,\,1)^{T}\,,
\ee
and then $\bB$
\bel{id3}
\bB = (z-y,\,x-z,\,y-x)^{T}\,.
\ee
Note that $|\bB| =0$ only on the straight line $x=y=z$.  Hence $|\bB|>0$  on any smooth,
finite domain $\Omega$ that avoids this line, which is enough to achieve our goal.

\section{Conclusion}\label{con}

These results raise curious questions regarding the nature of 3D Euler flow with a passive scalar. Physically $\theta$
could represent, for instance, the concentration of a dye or a quantity of fine dust added to an Euler flow. As a passive
quantity it would be appear to be innocuous, but its presence introduces the gradient $\nabla\theta$ which interacts
with $\bom$ and thereby introduces the second passive quantity $q = \bom\cdot\nabla\theta$ into the dynamics. The
key result is the stretching relation for $\bB$ in (\ref{Bev1}), where $\bB$ is simply a vector tangent to the curve that
intersects the two material surfaces for $\theta$ and $q$. The first null point in $|\bB|$ then drives
$\int_{0}^{t}\|\mathcal{S}\|_{L^{\infty}(\Omega)}\,d\tau \to\infty$, through the logarithmic singularity. The presence
of $\theta$ is therefore essential to this mechanism. This raises the question whether this singularity is of a fundamentally
different type than those that have been conjectured to develop in bare 3D Euler flow with no additional passive scalar?
\par\smallskip
The proof of Theorem \ref{Bthm} shows that it is essential that $|\bB|$ starts with no null points. This rules out the use
of periodic boundary conditions because, under these conditions, $\bB = \nabla q\times\nabla\theta$ has zeros for every
value of $t$. This is because   $\nabla\theta$ must have a null point in any periodic domain, whenever $\theta \in C^1$.  
Hence a comparison with the standard body of numerical experiments is not possible, although it would suggest that a 
numerical examination of the occurrence and nature of null points might be fruitful for flows in smooth bounded domains 
with the boundary conditions used in this paper.
\par\smallskip
A further variation on this problem is that of the 3D Euler equations with buoyancy, i.e., the inviscid, without diffusion, 
Bousinessq system, which can be written in the following dimensionless form
\bel{eul2a}
\frac{D\bu}{Dt} + \Theta \bk = - \nabla p\,,\qquad\qquad \frac{D\Theta}{Dt} = 0\,,\qquad
\qquad\qquad\mbox{div}\,\bu = 0\,.
\ee
$\Theta$ is a dimensionless temperature and appears because the fluid density has been taken to be proportional to $\Theta$
in the Boussinesq approximation. This adds an extra $\nabla\Theta\times\bk$ term to the left hand side of equation (\ref{eulvort1}).
However, this extra term makes no contribution to equation (\ref{q1B}) and so $q$ remains passive. The BKM criterion for system 
(\ref{eul2a}), on a finite smooth domain $\Omega$, can be  re-worked without much difficulty by following the work Ferrari (1993 
and Shirota and Yanagisawa (1993).
\par\medskip\noindent
\textbf{Acknowledgement\,:} The authors would like to thank a referee for some very helpful comments. John Gibbon also thanks 
the Isaac Newton Institute for Mathematical Sciences
Cambridge for its hospitality (July -- December 2012) on the programme ``Topological Dynamics in the Physical and
Biological Sciences'' under whose auspices this work was done. He would also  like to thank Darryl Holm of Imperial
College London for several discussions on this problem.  The work of E. S. Titi is supported in part by the Minerva
Stiftung/Foundation and also by the NSF grants DMS-1009950, DMS-1109640 and DMS-1109645.


\end{document}